\documentclass[runningheads]{llncs}
\usepackage[T1]{fontenc}
\usepackage{graphicx,verbatim}
\usepackage{color}
\usepackage{booktabs}
\usepackage{hyperref}
\usepackage{amsmath, amssymb}
\usepackage{cite}
\usepackage{orcidlink}

\begin{document}

\title{SV-DRR: High-Fidelity Novel View X-Ray Synthesis Using Diffusion Model}

\author{Chun Xie\inst{1}\orcidlink{0000-0003-4936-7404} \and
Yuichi Yoshii\inst{2}\orcidlink{0000-0003-1447-4664} \and
Itaru Kitahara\inst{1}\orcidlink{0000-0002-5186-789X}}
\authorrunning{C. Xie et al.}
\institute{
Center for Computational Sciences, University of Tsukuba, Japan 
\email{\{xiechun,kitahara\}@ccs.tsukuba.ac.jp} \and
Tokyo Medical University Ibaraki Medical Center, Japan
\email{yyoshii@tokyo-med.ac.jp}
}

\maketitle              % typeset the header of the contribution
\begin{abstract}
X-ray imaging is a rapid and cost-effective tool for visualizing internal human anatomy. While multi-view X-ray imaging provides complementary information that enhances diagnosis, intervention, and education, acquiring images from multiple angles increases radiation exposure and complicates clinical workflows. To address these challenges, we propose a novel view-conditioned diffusion model for synthesizing multi-view X-ray images from a single view. Unlike prior methods, which are limited in angular range, resolution, and image quality, our approach leverages the Diffusion Transformer to preserve fine details and employs a weak-to-strong training strategy for stable high-resolution image generation. Experimental results demonstrate that our method generates higher-resolution outputs with improved control over viewing angles. This capability has significant implications not only for clinical applications but also for medical education and data extension, enabling the creation of diverse, high-quality datasets for training and analysis. Our code is available at \url{https://github.com/xiechun298/SV-DRR}.

% Authors must provide keywords and are not allowed to remove this Keyword section.

\end{abstract}
\begin{figure}[tb]
\includegraphics[width=\textwidth]{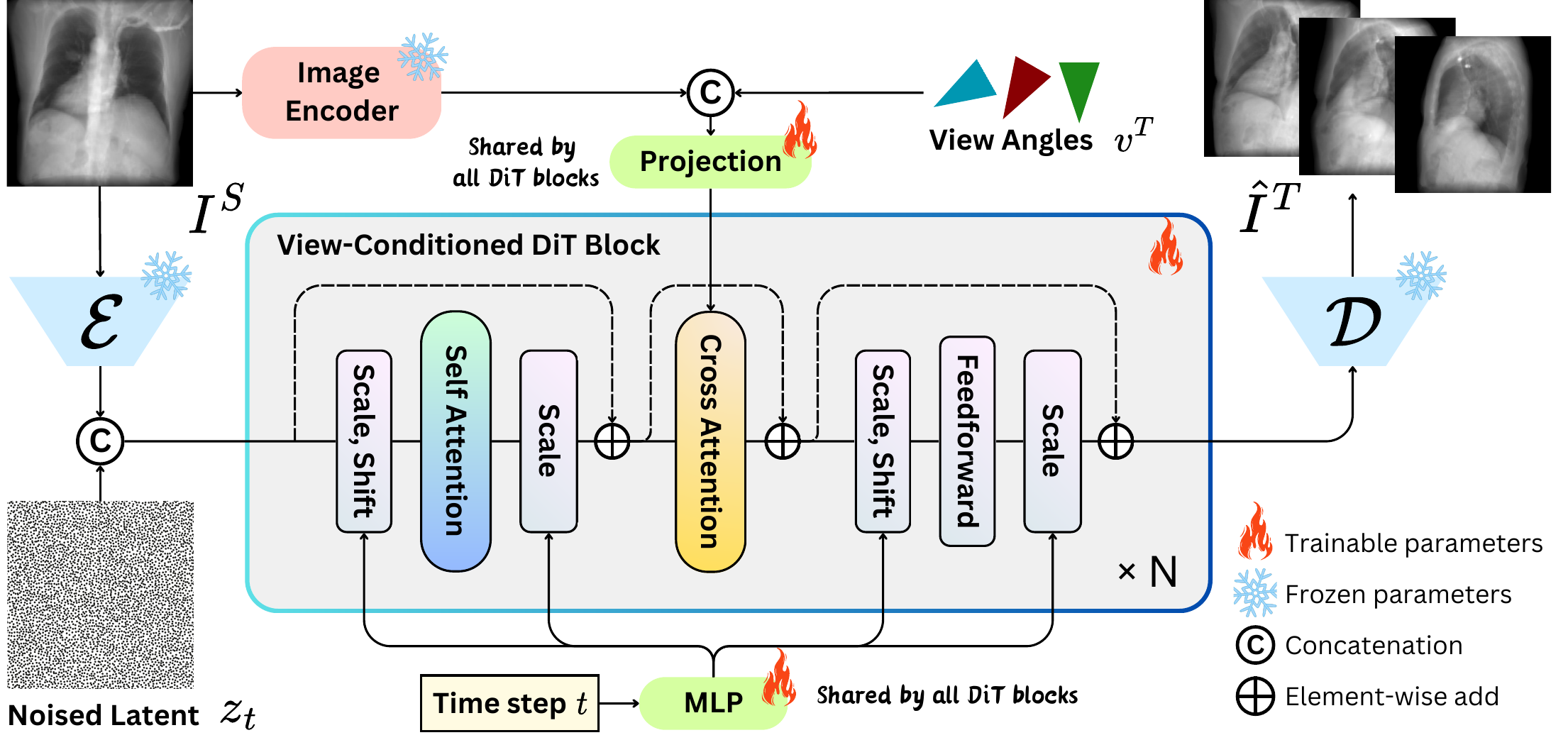}
\caption{Overview of SV-DRR. Given a source X-ray image \( I^S \) and relative target views \( v^T \), SV-DRR synthesizes realistic X-ray projections \( \hat{I}^T \) corresponding to \( v^T \) from Gaussian noise \( z_t \) via a denoising DiT. The denoising process takes place in the latent space of a VAE, where $\mathcal{E}$, and $\mathcal{D}$ denote the encoder and decoder, respectively.} \label{fig:overview}
\end{figure}

\section{Introduction}
X-ray imaging is a widely used, cost-effective technique for visualizing internal human anatomy. It utilizes high-energy X-rays to penetrate the body and capture residual radiation on a flat detector, enabling non-invasive diagnosis. However, conventional radiography is typically limited to a single-view projection, such as in chest radiography, where only a frontal image is acquired per session. While radiologists can infer 3D spatial relationships from these 2D images, this process relies heavily on subjective intuition, which is challenging to standardize and visualize explicitly. Furthermore, acquiring additional views requires multiple exposures, increasing radiation risk and complicating clinical workflows.

To address these challenges, novel view synthesis offers a promising solution\cite{Peng2020XraySyn,SHEN2022102372,MedNeRF} by generating multiple perspectives from a single X-ray image. This capability could facilitate more intuitive assessments, particularly in settings without access to CT imaging, reduce radiation exposure, and mitigate the effects of patient movement. Moreover, novel view synthesis is valuable for applications such as sparse-view CT reconstruction\cite{Kyung2023PerX2CT, Jeong2024DX2CTDM}, improving imaging efficiency and expanding the utility of X-ray imaging in diagnostic and educational settings.

This paper presents SV-DRR(Fig.\ref{fig:overview}), a novel view-conditioned diffusion model for multi-view X-ray synthesis, overcoming the limitations of existing methods in angular range, resolution, and stability. The name SV-DRR, short for Single-View DRR, is inspired by Digitally Reconstructed Radiography (DRR). Unlike DRR, which renders X-ray projections from a 3D CT volume, our method synthesizes novel views directly from a single 2D projection. Our approach formulates view synthesis as a view-conditioned image variation task, leveraging a Diffusion Transformer (DiT)\cite{peebles2023DiT} to enhance anatomical coherence and detail preservation. Additionally, a weak-to-strong training strategy stabilizes high-resolution synthesis, while a densely sampled dataset with view annotations provides a robust foundation for training and evaluation across varying viewpoints.

Our main contributions are summarized as follows:
\begin{itemize}
\item \textbf{\emph{Advancing View Synthesis in X-ray Imaging:}} We propose a view-conditioned DiT model that overcomes previous limitations in angular coverage and resolution, significantly expanding the range of synthesizable viewpoints. This enables more flexible and clinically useful multi-view X-ray synthesis.

\item \textbf{\emph{Improving Stability and Resolution:}} Our weak-to-strong training strategy enhances synthesis quality, allowing for high-resolution X-ray image generation with improved stability and computational efficiency, making it more feasible for real-world applications.

\item \textbf{\emph{Enhancing Data Availability for Training and Evaluation:}} We create a densely sampled dataset with precise view annotations, enabling robust training for synthesizing multi-view X-ray images and facilitating research in sparse-view CT reconstruction and other related fields.
\end{itemize}

\section{Related Work}\label{sec:rw}

\textbf{GAN-based methods}: GAN\cite{GAN}-based methods have been widely explored for single-image view synthesis, particularly in natural image domains, where they infer missing structures to generate novel views\cite{Park2017transformation}. In X-ray imaging, recent approaches leverage CT priors and geometric constraints \cite{Peng2020XraySyn, SHEN2022102372, MedNeRF} to enhance anatomical fidelity and structural coherence. However, these methods remain limited in view range and resolution, restricting their applicability for high-quality, multi-angle X-ray synthesis.

\textbf{Diffusion-based methods}: Recently, diffusion models have demonstrated exceptional performance in image generation by iteratively refining noisy inputs. Foundational works such as Denoising Diffusion Probabilistic Models (DDPM) \cite{ho2020ddpm} and Denoising Diffusion Implicit Models (DDIM) \cite{song2021ddim} established the basis for modern diffusion-based synthesis. These advancements have led to the development of powerful conditional image generation models \cite{ rombach2022stablediffusion, Podell2023SDXLIL, chen2023pixartalpha, chen2024pixartsigma, Siamese-Diffusion}. More recently, diffusion models have been applied to novel view synthesis \cite{Rockwell2021PixelSynth, liu2023zero1to3, watson2023novel, chan2023genvs}, where camera parameters serve as conditioning inputs for generating new viewpoints. Building on these developments, we propose a view-conditioned diffusion model for X-ray synthesis, which incorporates explicit view embeddings and latent-space conditioning to achieve high-quality multi-view generation.

\section{Proposed Method}\label{sec:method}
\subsection{Overview}
X-ray view synthesis aims to generate novel projections from a given input X-ray image and its associated view parameters. Given an input source image $I^S$ with corresponding view parameters $v^S$, the objective is to synthesize an X-ray image  $\hat{I}^T$ from a target viewpoint defined by $v^T$. By modeling the structural relationships between different projections, our method estimates the conditional distribution:
\begin{equation}
\hat{I}^T \sim \mathcal{P}(I | I^S, v^S, v^T),
\end{equation}
ensuring that the generated images accurately preserve anatomical structures of $I^S$.

\subsection{View-Conditioned Diffusion Transformer}  

Our View-Conditioned Diffusion Transformer (VCDiT) follows the Latent Diffusion Model (LDM) framework \cite{Rombach_2022_CVPR_LDM}, where the diffusion process is performed in VAE latent space \cite{kingma2019vae}, reducing computational complexity while preserving fine details. As shown in Fig.\ref{fig:overview}, the denoising network \( \epsilon_\theta \) is based on DiT, enhanced with a cross-attention mechanism inspired by \cite{chen2023pixartalpha} to inject view-conditioning. To improve efficiency, we use a shared AdaLN-Single layer for timestep embedding \( t \) and a shared learnable linear layer (projection) to efficiently map the view-conditioning signal to the VCDiT latent space.

To enable view-conditioned synthesis, VCDiT integrates conditioning through two complementary streams. First, the embedding of the source image \( I^S \) is concatenated with an encoded relative polar coordinate between the target view \( v^T \) and the source view \( v^S \), forming a view embedding that provides spatial transformation information for controlled generation. Second, the source image latent is channel-concatenated with the noised target latent before denoising, reinforcing structural alignment and improving synthesis quality.

The training objective follows standard conditional diffusion models, minimizing the loss:
\begin{equation}
\mathcal{L} = \mathbb{E}_{z_0, c, \epsilon, t} \|\epsilon - \epsilon_\theta(z_t, c(I^S,v^S,v^T), t)\|^2_2
\end{equation}
where \( \epsilon \sim \mathcal{N}(0, I) \) represents Gaussian noise added to the latent, and \( z_0 \) and \( z_t \) denote the original and noised latents of the target image \( I^T \), respectively. The function \( c \) represents the conditioning mechanism.

By incorporating these conditioning mechanisms, VCDiT enables the synthesis of high-resolution novel X-ray views while maintaining anatomical accuracy and spatial coherence.

% $z_0 = \mathcal{E}(I^s) z'_t=z_t \oplus_c \mathcal{E}(I^t) $

\subsection{Weak-to-Strong Training Strategy} 
Our training strategy follows a progressive resolution refinement approach, where the model is initially trained on lower-resolution(LR) images before gradually increasing the resolution. This method allows the model to efficiently learn fundamental X-ray image features at a reduced computational cost before adapting to high-resolution(HR) image synthesis.

Previous studies have reported a decline in performance during the transition from LR to HR training due to inconsistencies in positional embeddings across different resolutions. To address this issue, we follow the approach of \cite{Podell2023SDXLIL,chen2023pixartalpha,chen2024pixartsigma} and adopt the positional embedding interpolation trick described in \cite{Xie_2023_ICCV}, where the HR model’s positional embeddings are initialized by interpolating those from the LR model. This approach mitigates discrepancies between resolutions, significantly improving the stability and efficiency of fine-tuning at HR X-ray images while maintaining structural coherence and detail accuracy.

\section{Experiment and Analysis}\label{sec:exp} 

To validate the superiority of SV-DRR, we compare it with multiple state-of-the-art view synthesis approaches on simulated X-ray images generated from CT volumes. The experiments assess model performance across different resolutions and view sets, evaluating both fidelity and robustness in synthesizing novel X-ray views.

\subsection{Dataset}
\textbf{\emph{LIDC-IDRI-DRR}} The Lung Image Database Consortium image collection (LIDC-IDRI)\cite{lidcidri}  consists of 1,012 chest CT scans. To ensure higher image fidelity and better reconstruction quality, we exclude CT scans with a slice thickness greater than 2.5 mm, resulting in a total of 889 CT volumes. We randomly select 16 volumes for evaluation and use the rest for training. The dataset will be released along with our code.

For preprocessing, we use 3D Slicer\cite{3dslicerweb} to remove CT tables. Chest X-ray images are synthesized from 1,500 views for each CT scan using DiffDRR\cite{diffdrr} to capture a wide range of perspectives, enhancing the robustness of our training dataset. We render X-ray images separately for different resolutions to avoid any loss in image quality caused by resizing. The view orientations are directed toward the center of the CT volumes. View positions are sampled on a hemisphere with a radius of 1.8 meters using Fibonacci lattice sampling. The first view is set to the standard Posterior-Anterior (PA) perspective and is always used as the source image in our experiment. 

\subsection{Implementation Details}
For latent encoding, we employ the VAE\cite{kingma2019vae} used in SDXL\cite{Podell2023SDXLIL}. We utilize CLIP\cite{Radford2021CLIP} as the image encoder to generate image embeddings, which serve as conditioning inputs for the VCDiT model's cross-attention mechanism, ensuring improved feature alignment and synthesis quality. Additionally, our VCDiT model is initialized using PixArt-$\mathrm{\Sigma}$-256\cite{chen2024pixartsigma} weights, leveraging its robust feature extraction capabilities.

Our model is trained using the AdamW optimizer with the learning rates set to $5\times10^{-6}$ for 256 resolution, $3\times10^{-6}$ for 512 resolution, and $1\times10^{-6}$ for 1024 resolution. The batch sizes are 64 for 256, 32 for 512, and 8 for 1024. Training is carried out on a single H100 GPU: 200K steps for 256 resolution, and 100K steps for 512 and 1024 resolutions. For sampling, we use DPMSolver\cite{dpmsolver} with an inference step of 20 and a guidance scale of 3.

\subsection{Comparison with State-of-the-art Methods} 
Table\ref{tab:experiment} presents a quantitative comparison of our method against state-of-the-art methods, including XraySyn\cite{Peng2020XraySyn}, Zero123\cite{liu2023zero1to3}, and Zero123-XL\cite{objaverseXL}, using PSNR, SSIM\cite{wang2004image}, LPIPS\cite{zhang2018unreasonable}, and FID\cite{heusel2017gans} metrics.

We evaluate performance using two distinct view sets. The first set \textit{Simple} involves varying the azimuth angle from -90° to 90° in 5° increments, resulting in 36 novel views. This setup resembles typical chest CT rotations and aligns with XraySyn’s testing conditions. The second set \textit{Hemisphere} is more challenging, consisting of 1,499 views distributed on a hemisphere with positions sampled using Fibonacci lattice sampling. This setup assesses the flexibility of each method, which is crucial for synthesizing X-ray images of other body parts like the knee, shoulder, or pelvis.

Our method achieves superior performance across all evaluation metrics in both view sets, demonstrating its capability to synthesize high-fidelity novel X-ray views even for large angular displacements. This improvement is driven by our densely sampled LIDC-IDRI-DRR dataset, which provides diverse training data, and our VCDiT model, which effectively captures the relationship between view conditioning and the preservation of fine details as well as structural coherence in synthesized images.

Furthermore, the minimal variance in evaluation metrics across different output resolutions highlights the stability of our approach in generating high-resolution X-ray images. This robustness is attributed to our weak-to-strong training strategy, which ensures consistency and reliability in high-resolution synthesis.

\begin{table}[tb]
\centering
\caption{Quantitative comparison of novel view synthesis on LIDC-IDRI-DRR}
\label{tab:experiment}
\begin{tabular}{@{}lcccccccc@{}} \toprule
& \multicolumn{4}{c}{Simple views} & \multicolumn{4}{c}{Hemisphere views}
\\\cmidrule(lr){2-5}\cmidrule(lr){6-9}
                                & SSIM$\uparrow$    & PSNR$\uparrow$    & LPIPS$\downarrow$ & FID$\downarrow$   & SSIM$\uparrow$    & PSNR$\uparrow$    & LPIPS$\downarrow$ & FID$\downarrow$    \\ \midrule
XraySyn \cite{Peng2020XraySyn}  & 0.4634            & 8.7670            & 0.4671            & 1.1902            & 0.2226            & 3.4804            & 0.2484            & 0.8450 \\
Zero123 \cite{liu2023zero1to3}  & 0.2933            & 8.2895            & 0.4849            & 0.9060            & 0.1217            & 3.6117            & 0.2714            & 1.0777 \\      
Zero123-XL \cite{objaverseXL}   & 0.5092            & 13.3156           & 0.3205            & 0.5412            & 0.2146            & 5.6295            & 0.1941            & 0.6832 \\
\cmidrule(lr){1-9}
SV-DRR 256                 & 0.7374            & 22.5136           & 0.1170            & \textbf{0.1880}   & 0.3600            & 10.8452           & 0.0640            & \textbf{0.1493} \\
SV-DRR 512                 & \textbf{0.7509}   & \textbf{23.3984}  & \textbf{0.1073}   & 0.2040            & \textbf{0.3680}   & \textbf{11.2855}  & \textbf{0.0588}   & 0.1693 \\
SV-DRR 1024                & 0.7336            & 22.7290           & 0.1191            & 0.1955            & 0.3600            & 10.9678           & 0.0644            & 0.1753 \\ 
\bottomrule
\end{tabular}
\end{table}

\subsection{Visualization} 
\begin{figure}[htbp]
\includegraphics[width=0.95\textwidth]{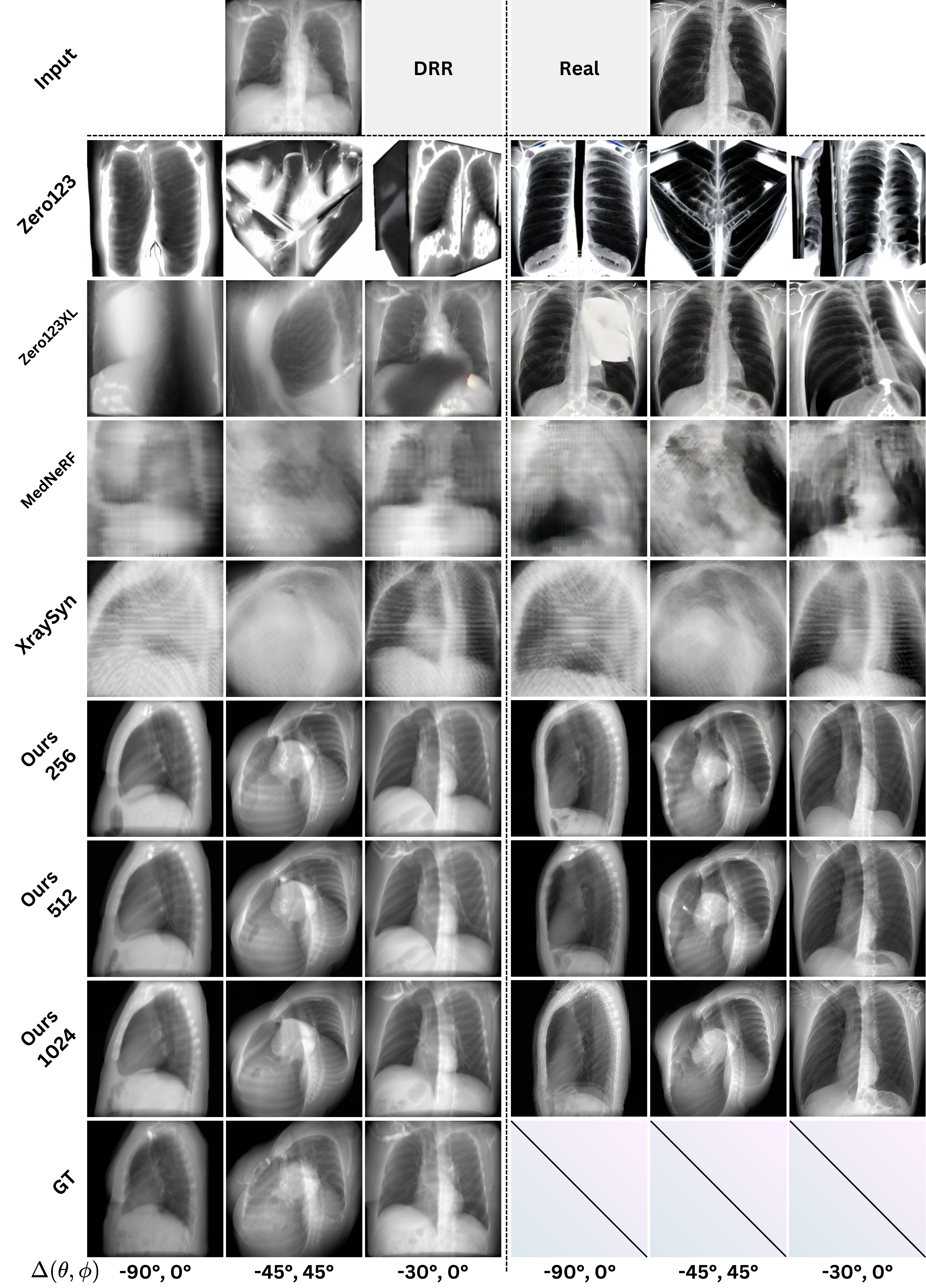}
\caption{Comparison of synthesized X-ray images at different resolutions using our method (256, 512, 1024) against baselines. Our method achieves superior fidelity in structure and orientation for both simulated DRR(left) and real X-ray(right) input. $\theta$ and $\phi$ represent the azimuthal and elevation angles of the synthesized view, respectively. Our method's different resolutions are normalized for better visualization. Additionally, the brightness and contrast of XraySyn images were manually adjusted for better comparison without introducing any negative effects on the results.} \label{fig:visulizetion}
\end{figure}

Fig.\ref{fig:visulizetion} presents a comparative analysis of view synthesis results generated by our method at resolutions of 256, 512, and 1024, alongside Zero123\cite{liu2023zero1to3}, Zero123-XL\cite{objaverseXL}, XraySyn\cite{Peng2020XraySyn}, and MedNeRF\cite{MedNeRF}, as well as the ground truth images from LIDC-IDRI-DRR. 

We evaluate our method using both simulated and real X-ray images as inputs. Across all tested resolutions, our approach consistently outperforms state-of-the-art methods, demonstrating superior fidelity in preserving anatomical structures and maintaining spatial consistency. XraySyn and MedNeRF retain the main anatomical structures only within a limited angular displacement, with increasing noise at larger angles. Zero123 produces transformations that resemble rigid rotations of a cubic object, leading to unrealistic projections. Interestingly, Zero123-XL, despite its enhanced zero-shot learning capabilities, struggles to incorporate the given conditions effectively, resulting in random distortions of the input image.

% Additional visualization results are provided in the supplementary materials for further qualitative evaluation.

\section{Subjective Evaluation of Image Realism}

\subsection{Experimental Design}
To evaluate the realism of the generated chest X-ray images, we conducted a user study involving medical experts. The objective was to determine whether experts could reliably distinguish between two sets of generated images: one produced using our proposed model and the other synthesized by applying DiffDRR to the LIDC-IDRI dataset to simulate real multi-view X-ray images. 

Fifteen board-certified medical experts and clinical practitioners evaluated 50 pairs of images. Each pair consisted of one image generated using DiffDRR and one image generated by our method, both from the same view angle. Experts were asked to determine which image in each pair appeared more real. The classification was conducted using a web-based interface, and each expert completed the evaluation independently.

\subsection{Results}

The average classification accuracy across participants was \textbf{48.7\%} (range: \textbf{40.0\%--58.0\%}), indicating performance close to random chance (50\%). A \textbf{one-sample t-test} against the null hypothesis of 50\% accuracy yielded a \textbf{t-statistic of $-1.000$} with a \textbf{p-value of 0.334}, showing no significant difference from chance. A \textbf{binomial test} on the total correct classifications (\textbf{365 out of 750}) produced a \textbf{p-value of 0.413}, further supporting this conclusion.

% \subsection{Discussion} 
The results strongly suggest that the images generated by our method are indistinguishable from those simulated using DiffDRR, even by medical experts. The absence of statistical significance in classification performance implies that the generated images possess high realism and fidelity. These findings highlight the potential of our method for data augmentation, simulation, and training applications in radiology.

\section{Conclusion}\label{sec:con} 
We introduced SV-DRR, a novel diffusion-based X-ray view synthesis method, addressing the limitations of previous works in terms of angular range, resolution, and stability. Using DiT and a weak-to-strong training strategy, our method generates high-resolution, anatomically coherent X-ray images. Additionally, our densely sampled dataset with view annotations enhances training and evaluation. Experimental results demonstrate that our method surpasses state-of-the-art approaches in quality and flexibility. Future work will focus on improving cross-view consistency to enhance anatomical alignment and realism, further advancing the applicability of multi-view X-ray synthesis in medical imaging.

% \begin{credits}
% \subsubsection{\ackname} 
% This work was supported in part by JSPS KAKENHI (grant number JP25100497), AMED (grant number JP250126803), and the Multidisciplinary Cooperative Research Program in CCS, University of Tsukuba.

% \subsubsection{\discintname}
% The authors have no competing interests to declare that are
% relevant to the content of this article.
% \end{credits}

\bibliographystyle{splncs04}
\bibliography{paper}

\begin{thebibliography}{10}
\providecommand{\url}[1]{\texttt{#1}}
\providecommand{\urlprefix}{URL }
\providecommand{\doi}[1]{https://doi.org/#1}

\bibitem{3dslicerweb}
3d slicer. \url{https://www.slicer.org/}, accessed: 2024-10-18

\bibitem{lidcidri}
Armato~III, S.G., McLennan, G., Bidaut, L., McNitt-Gray, M.F., Meyer, C.R., Reeves, A.P., Zhao, B., Aberle, D.R., Henschke, C.I., Hoffman, E.A., et~al.: The lung image database consortium (lidc) and image database resource initiative (idri): a completed reference database of lung nodules on ct scans. Medical physics  \textbf{38}(2),  915--931 (2011)

\bibitem{chan2023genvs}
Chan, E.R., Nagano, K., Chan, M.A., Bergman, A.W., Park, J.J., Levy, A., Aittala, M., De~Mello, S., Karras, T., Wetzstein, G.: {GeNVS}: Generative novel view synthesis with {3D}-aware diffusion models. In: Proceedings of the IEEE/CVF International Conference on Computer Vision (ICCV) (2023)

\bibitem{chen2024pixartsigma}
Chen, J., Ge, C., Xie, E., Wu, Y., Yao, L., Ren, X., Wang, Z., Luo, P., Lu, H., Li, Z.: Pixart-$\mathrm{\Sigma}$: Weak-to-strong training of diffusion transformer for 4k text-to-image generation (2024)

\bibitem{chen2023pixartalpha}
Chen, J., Yu, J., Ge, C., Yao, L., Xie, E., Wu, Y., Wang, Z., Kwok, J., Luo, P., Lu, H., Li, Z.: Pixart-$\alpha$: Fast training of diffusion transformer for photorealistic text-to-image synthesis (2023)

\bibitem{MedNeRF}
Corona-Figueroa, A., Frawley, J., Taylor, S.B., Bethapudi, S., Shum, H.P.H., Willcocks, C.G.: Mednerf: Medical neural radiance fields for reconstructing 3d-aware ct-projections from a single x-ray. In: 2022 44th Annual International Conference of the IEEE Engineering in Medicine \& Biology Society (EMBC). pp. 3843--3848 (2022). \doi{10.1109/EMBC48229.2022.9871757}

\bibitem{objaverseXL}
Deitke, M., Liu, R., Wallingford, M., Ngo, H., Michel, O., Kusupati, A., Fan, A., Laforte, C., Voleti, V., Gadre, S.Y., VanderBilt, E., Kembhavi, A., Vondrick, C., Gkioxari, G., Ehsani, K., Schmidt, L., Farhadi, A.: Objaverse-xl: A universe of 10m+ 3d objects. arXiv preprint arXiv:2307.05663  (2023)

\bibitem{GAN}
Goodfellow, I., Pouget-Abadie, J., Mirza, M., Xu, B., Warde-Farley, D., Ozair, S., Courville, A., Bengio, Y.: Generative adversarial nets. In: Ghahramani, Z., Welling, M., Cortes, C., Lawrence, N., Weinberger, K. (eds.) Advances in Neural Information Processing Systems. vol.~27. Curran Associates, Inc. (2014)

\bibitem{diffdrr}
Gopalakrishnan, V., Golland, P.: Fast auto-differentiable digitally reconstructed radiographs for solving inverse problems in intraoperative imaging. In: Workshop on Clinical Image-Based Procedures. pp. 1--11. Springer (2022)

\bibitem{heusel2017gans}
Heusel, M., Ramsauer, H., Unterthiner, T., Nessler, B., Hochreiter, S.: Gans trained by a two time-scale update rule converge to a local nash equilibrium. In: Advances in Neural Information Processing Systems. pp. 6626--6637 (2017)

\bibitem{ho2020ddpm}
Ho, J., Jain, A., Abbeel, P.: Denoising diffusion probabilistic models. In: Advances in Neural Information Processing Systems. vol.~33, pp. 6840--6851. Curran Associates, Inc. (2020)

\bibitem{Jeong2024DX2CTDM}
Jeong, Y.S., Yoo, H.B., Chun, I.Y.: Dx2ct: Diffusion model for 3d ct reconstruction from bi or mono-planar 2d x-ray(s). ArXiv  \textbf{abs/2409.08850} (2024)

\bibitem{kingma2019vae}
Kingma, D.P., Welling, M.: An introduction to variational autoencoders. Foundations and Trends in Machine Learning  \textbf{12}(4),  307--392 (2019). \doi{10.1561/2200000056}

\bibitem{Kyung2023PerX2CT}
Kyung, D., Jo, K., Choo, J., Lee, J., Choi, E.: Perspective projection-based 3d ct reconstruction from biplanar x-rays. ICASSP 2023 - 2023 IEEE International Conference on Acoustics, Speech and Signal Processing (ICASSP) pp.~1--5 (2023)

\bibitem{liu2023zero1to3}
Liu, R., Wu, R., Hoorick, B.V., Tokmakov, P., Zakharov, S., Vondrick, C.: Zero-1-to-3: Zero-shot one image to 3d object (2023)

\bibitem{dpmsolver}
Lu, C., Zhou, Y., Bao, F., Chen, J., LI, C., Zhu, J.: Dpm-solver: A fast ode solver for diffusion probabilistic model sampling in around 10 steps. In: Koyejo, S., Mohamed, S., Agarwal, A., Belgrave, D., Cho, K., Oh, A. (eds.) Advances in Neural Information Processing Systems. vol.~35, pp. 5775--5787. Curran Associates, Inc. (2022)

\bibitem{Park2017transformation}
Park, E., Yang, J., Yumer, E., Ceylan, D., Berg, A.C.: Transformation-grounded image generation network for novel 3d view synthesis. In: 2017 IEEE Conference on Computer Vision and Pattern Recognition (CVPR). pp. 702--711 (2017). \doi{10.1109/CVPR.2017.82}

\bibitem{peebles2023DiT}
Peebles, W., Xie, S.: Scalable diffusion models with transformers. In: Proceedings of the IEEE/CVF International Conference on Computer Vision (ICCV). pp. 4195--4205 (2023). \doi{10.1109/ICCV53048.2023.00420}

\bibitem{Peng2020XraySyn}
Peng, C., Liao, H., Wong, G.G., Luo, J., Zhou, S.K., Chellappa, R.: Xraysyn: Realistic view synthesis from a single radiograph through ct priors. In: AAAI Conference on Artificial Intelligence (12 2020). \doi{10.1609/aaai.v35i1.16120}

\bibitem{Podell2023SDXLIL}
Podell, D., English, Z., Lacey, K., Blattmann, A., Dockhorn, T., Müller, J., Penna, J., Rombach, R.: Sdxl: Improving latent diffusion models for high-resolution image synthesis (2023), \url{https://arxiv.org/abs/2307.01952}

\bibitem{Siamese-Diffusion}
Qiu, K., Gao, Z., Zhou, Z., Sun, M., Guo, Y.: Noise-consistent siamese-diffusion for medical image synthesis and segmentation. In: 2017 IEEE Conference on Computer Vision and Pattern Recognition (CVPR) (2025). \doi{10.48550/arXiv.2505.06068}

\bibitem{Radford2021CLIP}
Radford, A., Kim, J.W., Hallacy, C., Ramesh, A., Goh, G., Agarwal, S., Sastry, G., Askell, A., Mishkin, P., Clark, J., Krueger, G., Sutskever, I.: Learning transferable visual models from natural language supervision. In: International Conference on Machine Learning (2021)

\bibitem{Rockwell2021PixelSynth}
Rockwell, C., Fouhey, D.F., Johnson, J.: {PixelSynth}: Generating a 3d-consistent experience from a single image. In: Proceedings of the IEEE/CVF International Conference on Computer Vision (ICCV) (2021)

\bibitem{rombach2022stablediffusion}
Rombach, R., Blattmann, A., Lorenz, D., Esser, P., Ommer, B.: High-resolution image synthesis with latent diffusion models. In: Proceedings of the IEEE/CVF Conference on Computer Vision and Pattern Recognition. pp. 10684--10695 (2022)

\bibitem{Rombach_2022_CVPR_LDM}
Rombach, R., Blattmann, A., Lorenz, D., Esser, P., Ommer, B.: High-resolution image synthesis with latent diffusion models. In: Proceedings of the IEEE/CVF Conference on Computer Vision and Pattern Recognition (CVPR). pp. 10684--10695 (June 2022)

\bibitem{SHEN2022102372}
Shen, L., Yu, L., Zhao, W., Pauly, J., Xing, L.: Novel-view x-ray projection synthesis through geometry-integrated deep learning. Medical Image Analysis  \textbf{77},  102372 (2022). \doi{10.1016/j.media.2022.102372}

\bibitem{song2021ddim}
Song, J., Meng, C., Ermon, S.: Denoising diffusion implicit models. In: International Conference on Learning Representations (2021)

\bibitem{wang2004image}
Wang, Z., Bovik, A.C., Sheikh, H.R., Simoncelli, E.P.: Image quality assessment: from error visibility to structural similarity. IEEE Transactions on Image Processing  \textbf{13}(4),  600--612 (2004)

\bibitem{watson2023novel}
Watson, D., Chan, W., Martin-Brualla, R., Ho, J., Tagliasacchi, A., Norouzi, M.: Novel view synthesis with diffusion models. In: International Conference on Learning Representations (2023)

\bibitem{Xie_2023_ICCV}
Xie, E., Yao, L., Shi, H., Liu, Z., Zhou, D., Liu, Z., Li, J., Li, Z.: Difffit: Unlocking transferability of large diffusion models via simple parameter-efficient fine-tuning. In: Proceedings of the IEEE/CVF International Conference on Computer Vision (ICCV). pp. 4230--4239 (October 2023). \doi{10.1109/ICCV51070.2023.00390}

\bibitem{zhang2018unreasonable}
Zhang, R., Isola, P., Efros, A.A., Shechtman, E., Wang, O.: The unreasonable effectiveness of deep features as a perceptual metric. In: Proceedings of the IEEE Conference on Computer Vision and Pattern Recognition. pp. 586--595 (2018)

\end{thebibliography}





% LNCS Homepage, \url{http://www.springer.com/lncs}, last accessed 2023/10/25
% \end{thebibliography}
\end{document}